# An Automated Contact Tracing Approach for Controlling Covid-19 Spread Based on Geolocation Data from Mobile Cellular Networks

Md. Tanvir Rahman[1], Risala T. Khan[2], Muhammad R. A. Khandaker[3], and Md. Sifat Ar Salan[4]

[1]Department of Information and Communication Technology, Mawlana Bhashani Science and Technology University, Tangail-1902, Bangladesh
[2]Institute of Information Technology, Jahangirnagar University, Savar, Dhaka-1342, Bangladesh
[3]School of Engineering and Physical Sciences, Heriot-Watt University, Edinburgh, UK
[4]Department of Statistics, Mawlana Bhashani Science and Technology University, Tangail-1902, Bangladesh

Corresponding author: M. T. Rahman (e-mail: sajal.it.ju@gmail.com).

**ABSTRACT** The coronavirus (COVID-19) has appeared as the greatest challenge due to its continuous structural evolution as well as the absence of proper antidotes for this particular virus. The virus mainly spreads and replicates itself among mass people through close contact which unfortunately can happen in many unpredictable ways. Therefore, to slow down the spread of this novel virus, the only relevant initiatives are to maintain social distance, perform contact tracing, use proper safety gears, and impose quarantine measures. But despite being theoretically possible, these approaches are very difficult to uphold in densely populated countries and areas. Therefore, to control the virus spread, researchers and authorities are considering the use of smartphone based mobile applications (apps) to identify the likely infected persons as well as the highly risky zones to maintain isolation and lockdown measures. However, these methods heavily depend on advanced technological features and expose significant privacy loopholes. In this paper, we propose a new method for COVID-19 contact tracing based on mobile phone users' geolocation data. The proposed method will help the authorities to identify the number of probable infected persons without using smartphone based mobile applications. In addition, the proposed method can help people take the vital decision of when to seek medical assistance by letting them know whether they are already in the list of exposed persons. Numerical examples demonstrate that the proposed method can significantly outperform the smartphone app-based solutions.

**INDEX TERMS** Coronavirus, Covid-19, Contact Tracing, Geolocation, Pandemic.

## I. INTRODUCTION

Since World War II, the coronavirus (COVID-19) is being considered to be the most life-threatening event that happened in the history of mankind. It is a new family of coronavirus that changes its RNA structure frequently. As the virus had spread all on a sudden and the spread rate could not be predicted in prior, only a few countries have been able to control the penetration of this virus. The main reason behind their successful control was the ability to trace the contacts of the infected persons promptly and put them into isolation [1]. To slow down the coronavirus spread, three measures have been proven effective: i) contact tracing, ii) social distancing, and iii) quarantine. In particular, contact tracing plays a leading role in containing the spread when the R-value is below 1.0. Since the outbreak began, different countries adopted different types of contract tracing strategies based on their socio-economic conditions. A vast majority of these methods operate through smartphone-based mobile apps.

According to the statistical portal, it has been found that in 2019 almost 5.07 billion people in the world are using mobile phones as a medium of communication and an estimated 67% of the world population has a mobile phone. The statistics also reports that in 2014 the smartphone users were around 38% which increased to 50% by the end of 2018 [2]. It has been reported that the number of smartphone users worldwide today surpasses three billion [2]. From this growth, it is expected that the number of smartphone users will grow by one billion in a period of five years [3], [4]. Therefore, contact tracing can be easily done using mobile phone apps but the challenge is how the collected data can be handled securely and what information should be collected [5]. In addition, people's willingness to install the apps as well as the availability of smartphones in all areas of a society is also a major challenge. According to Ferretti et al. [6], [7], if 60% of the population of a country install the contact tracing app, the spread rate will automatically slow down. After collecting the data, the next obvious question is how to warn people effectively. Many approaches are now under consideration such as contact tracing, narrowcasting, broadcasting, and so on to alert people regarding the possibility of being infected or tracing the likely infected person's contact trajectory [8]. Accurate contact tracing data is vital for providing timely exposure notices. For that purpose, users' personal information such as mobility, details of the persons the suspected user contacted, etc. has to be disclosed. The app can provide the data with reliable accuracy only if the collected information is sufficient. But more information leads to increased breach of privacy which is a major concern these days.

For a clear understanding of the privacy concerns related to smartphone app-based contact tracing, one needs to know how such apps operate in practice. A typical contact tracing app works as follows: the app should be installed on an individual's cell phone and the Bluetooth of the phone must always remain on. When two users, having the same app installed, reach in close proximity, the app exchanges a unique identifier using Bluetooth which is stored either in the phone storage or in a centralized database. If a person is found to be COVID-19 positive, his mobile is taken to collect all the mobile numbers that had so far been stored in this mobile, and then those persons are informed as soon as possible. However, the authors in [9] raised some valid questions regarding the method being used in the Bluetooth-based contact tracing app, such as, how the anonymity of a user be protected from the app provider, how to prevent the snooper to access individual's data through this app, what measure the authority will take in case of contacted persons after knowing their details and so on. Therefore, not only finding the infected or likely infected person is enough but how to inform them or what measures should be taken to confine them is also important in the pandemic situation. In addition, solutions that seem effective for some countries may not be appropriate for other countries due to different social norms.

To address the aforementioned challenges in smartphone app-based contact tracing of COVID-19, we propose the idea of using mobile phone users' geolocation data for contact tracing. The proposed model does not require Bluetooth/Wi-Fi/NFC enabled cell phones since the responsible authorities will collect the data directly from the network providers. The procedure is detailed in the following sections.

Meanwhile, the rest of the paper is organized as follows: Section II presents an extensive review of existing systems, Section III focuses on the details the proposed model, Section IV compares the proposed model with existing systems from different viewpoints, then, experimental results are presented in section V. Finally, conclusions are drawn in section VI.

## II. REVIEW OF EXISTING SYSTEMS

After the declaration of WHO as COVID-19 outbreak a public health emergency of international concern (PHEIC) on 30th January 2020 [10, 11] and a pandemic on 11th March 2020 [12], many countries developed different contact tracing apps to monitor and control the spread in the country. According to a CNN report, the Health Code app [13] that is being used in many parts of China works as follows: the app asks people about their symptom as well as their travel history, the possibility of being in contact with a COVID-19 positive patient, their workplaces, residential addresses, phone numbers, passport numbers, national identity number, etc. will be verified. After verification, a color code will be sent to the person's mobile phone named as "QR Code" whose color can be either red, green or amber. Users with red code will have to go under government quarantine or self-quarantine for 14 days, users with amber code will go to quarantine for 7 days but users with green code are considered to be risk-free. The major problem in this app is if a person intentionally provides wrong information regarding his travel information or symptom or being in close contact with a COVID-19 positive patient, then he will get a green code and will likely affect more people before being identified.

South Korea was among the first few countries that were affected by the novel coronavirus after China. Hence, in South Korea, the first confirmed cases were reported on 20th January 2020 [14]. Even though in the initial stage the number of affected people in South Korea was quite high but they didn't go for any lockdown or roadblock or aggressive immigration control strategy. Rather they used a trace, test, and treat strategy which proved very effective as within a short period the curve of newly confirmed cases, as well as deaths, had been flattened at around mid-March [15]. To control the spreading, the Government of South Korea sent all the travelers who came from abroad into self-quarantine. During the quarantine state, the travelers forcefully used a self-diagnosis app and updated their health status regularly so that the Government can trace whether there are any suspected symptoms [16]. At the same time, the government also sent all of the individuals who had direct contact with those travelers to self-quarantine as well and followed the same monitoring scheme. But the main problem in this tracing process arose when the collected data was shared among many authorities such as police, health insurance, central government agencies, health care professionals, health care associations, and others. Hence, it is a direct violation of data privacy law in South Korea [15]. But during the pandemic situation, such law can be relaxed and that was what South Korea did at that time. Besides, it was suggested that if such a situation happens again, aggregated data rather than an individual's can be shared to all the concerned sectors to control the misuse of data [15].

In March 2020, Singapore Ministry of Health first released a contact tracing app called "TraceTogether" and its BLE-based protocol BlueTrace [17] where the tracing will be done through mobile phones' Bluetooth technology. For tracking purpose, a person has to install the app, and Bluetooth must be turned on always. A unique token is generated in the person's mobile during installation. Whenever two persons are in close proximity their phone will exchange that token through Bluetooth and preserve that token number in the phone storage [18].

The Pan-European Privacy-Preserving Proximity Tracing (PEPP-PT) [19], a joint project between Germany, France and Italy, proposed a centralized data center based on Bluetooth low energy (BLE)-based tracing technique where any traveler travelling within EU countries can use the same app to let the government know his contacted persons' list without needing any additional contact tracing app.

There are several major problems in the BLE-based tracing systems following both centralized and decentralized approaches. First, if the person does not use a smartphone featuring Bluetooth connectivity, then there is no way he or she can be traced. We acknowledge that in a first world country like Singapore or Germany, most of the people use smartphones but not necessarily everyone keeps the Bluetooth connectivity enabled in his/her device at all time. Second, although Bluetooth technology is considered as a cheap, reliable, and low power consuming option, any malicious user may access the information stored in the mobile devices through the Bluetooth link [20, 21, 22]. In [21], Naveed et al. explained the security issues of using mobile Bluetooth technology for android devices and in [23] the authors showed the security issues in the iOS platform. Data privacy is thus a major concern in Bluetooth based apps. For any contact tracing app, the main goal is to inform the person that he or she has been exposed to the virus. However, with Bluetooth-based apps, even if the person has not been in touch with any infected person but within the Bluetooth range of the person, the user of the phone will be flagged as a suspected one and may be subject to social bullying. It may also create a mass confusion if the number of such false alarm increases significantly. On the other hand, if we consider the scenario of a developing country like Bangladesh, Nigeria etc., where coronavirus is creating phobia among people as medical facilities to support the massive number of patients are not adequate and many hospitals are not even treating patients with other problems due to the suspicion of coronavirus, this type of contract tracing app will create mass panic rather than mass awareness and people will face social harassment.

In general, two types of data can be collected by using these apps, one is "proximity" which will show with whom the user has been in contact and "location" which will show in which area the user has roamed [8]. Both may cause false positive alarms and false-negative alarms [8]. A false-positive alarm will happen if the person who is identified being in contact either by location information or contact information may just pass the area at that time and didn't even come in direct contact with the infected person. A false-negative alarm means the app is failing to identify any suspected person which may happen due to many reasons as we have discussed earlier.

Interestingly, many existing works suggest decentralized solutions because less sensitive data can then be shared with central authorities and thus the end user's privacy can be preserved [24, 25]. However, the risk of the decentralized solution compared to the centralized solution is still a trade-off and beyond the scope of this work.

While most of the contact tracing apps are now using the BLE-based tracing technique, some researchers are also thinking about storing user's mobility information and only hand over this information only when found COVID-19 positive [26]. In terms of data security and privacy, this is more secure as logging the location without the contact's stored information is less private.

In [27], the authors proposed a Bluetooth based contact tracing app "CONTAIN" where they showed that the identity of the user will be completely anonymous and at the same time, they experimented different timing of turning the Bluetooth on (random, decentralized, centralized) of the mobile to compare the results. They found in their experimental results that even if the Bluetooth is turned on, only when the user goes to any public gathering, the performance of the app is better than if it is turned on randomly during different times of the day.

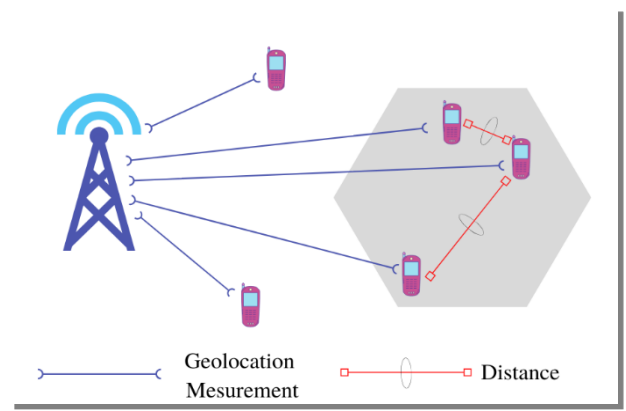

**FIGURE 1.** Contact tracing through cell phone network using geolocation data.

We have found that most of the COVID-19 contact tracing apps released so far are based on location/ proximity-based contact tracing, location-based hotspot reporting, and self-reported symptom tracking [17, 19, 24-33]. On the other hand, if the location data can be obtained from cellular mobile phone networks, then the issues related to the BLE-

3                                                                                                                                                                      VOLUME XX, 2017

based tracing technique can be eliminated as well as the technological requirements (e.g. phones with Bluetooth and smart features) can be waived. The authors of [34] have proposed an outdoor location tracking technique of mobile devices in a cellular network. They have shown that the proposed technique is nearly 88% accurate. In urban areas, the tracking performance is the best with median accuracies of up to 112m. It is also found that the cellular systems show promising positioning performance in indoor environments as well [35].

In this paper, we propose a contact tracing approach using cellular sim card geolocation data as opposed to the BLE-based tracing techniques. In the proposed model, a confirmed COVID19 patient's mobile number will be passed to the corresponding mobile operator to find its mobility information for the past 7 days. The operator will use the geolocation information to trace the mobility as shown in Fig.1.

In the proposed model, no Bluetooth/ Wi-Fi/ NFC enabled cell phone is needed as the operator will use the geolocation-based tracing approach by getting the location data directly from the base station to identify the likely infected persons. This method can avoid panic spread among people as it does not require continuous warning about their possibility of getting infected. The responsible authority can then identify the riskiest persons based on the intensity of contact with Covid-19 positive patients and instruct them to go for isolation/test if necessary. In addition, if a person feels to have developed any symptoms of COVID-19, the proposed method can also perform an initial screening by checking whether the cell phone number is enlisted in the list of suspects.

## III. PROPOSED CONTACT TRACING MODEL
In this section, we present the proposed cell phone geolocation-based contact tracing model. We structure the overall framework in three operational phases. Each phase has specific activities. The phases and their operations are discussed below.

### A. PHASE – I: Data Collection
In this phase, the primary data for the COVID-19 patients will be collected from the designated test centers. Then the infected areas will be shown using any map services (e.g. Google map). The activities to be performed at this stage are described below and the process is depicted in Fig. 2.

1) STEP 01
Whenever a person visits a COVID-19 test center, two basic information, such as the current address and the active mobile phone number(s) must be recorded in the local register of the test center.
2) STEP 02
If the person is found to be COVID-19 positive, then the previously recorded information (i.e. address and phone numbers) must be reported to the central database.
3) STEP 03
At this point, the number of infected persons in an individual area (e.g. Division, District, City, Road, etc.) will be calculated.
4) STEP 04
An application will be in place (e.g. iOS or Android App, or Web Application, etc.) where anyone can do a simple registration with the mobile number. The registered user can search for a particular location and get the number of infected persons around it via any map services (e.g. Google map).

### B. PHASE – II: Identifying Probable Occurrences
In the second phase, we focus on how the Government can identify the likely infected persons. The cell phone number of the Covid-19 positive patients can be traced to find out the possible occurrences as shown in Fig. 3. Here, any other cell phone user who came in near proximity (e.g. 2 meters) to the infected person during the last 7 days lies in the suspected list. But the mass people do not need to install any kind of application as the location data (latitude and longitude) will be collected from the cell phone network. Therefore, this process is not limited to the use of smart phones, rather we can get the data for any type of cell phone (e.g. feature phone) users. The step-by-step process is illustrated below, and the process is depicted in Fig. 4.
1) STEP 01
After getting the data (i.e. the list of infected persons) from the central database, each mobile number will be sent to the corresponding mobile operator to get the mobility information (latitude and longitude) of the respective cell phone users during the prior 7 days. At the same time, the operator will also be asked to provide the mobile numbers of the active cell phone users within those infected persons' mobility zones within a given distance.
2) STEP 02
The mobility zone information, acquired from the previous step, will then be sent to all other operators to identify all of the active cell phone users inside the zone at that time.

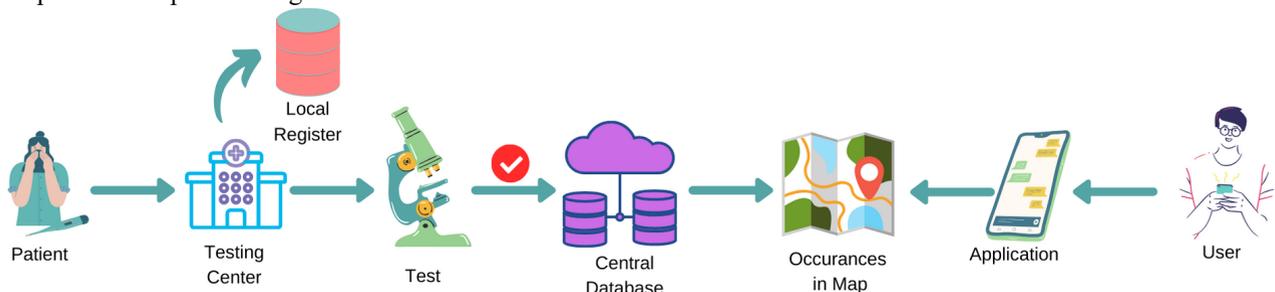

**FIGURE 2.** Flow diagram of Phase - I

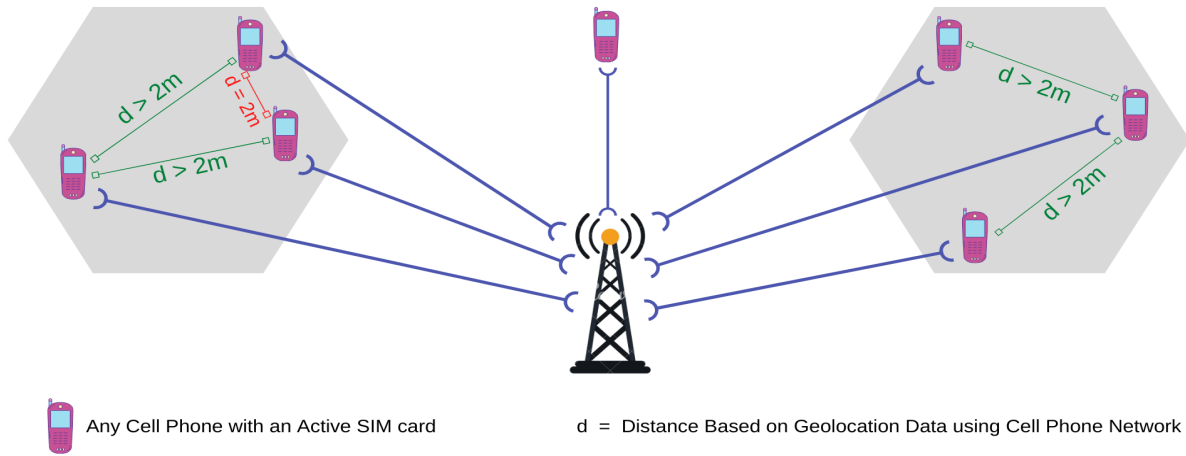

**FIGURE. 3.** Finding possible occurrences using geolocation data from mobile phone operators.

3) STEP 03
All the cell phone numbers (obtained from step 1 and 2) will be stored in the central database marked as "possible infected cell phone numbers".

4) STEP 04
Step 04: If a particular number is found to appear multiple times in the "possible infected cell phone numbers" list, then those cell phone users will be instructed to go for isolation/ COVID-19 test as appropriate.

### C. PHASE – III: User's Query

At this phase, if a person feels physically uncomfortable and wants to go for a test, the application can help to make the decision and predict the intensity. For this purpose, we propose activities as follows in stepwise and the process is depicted in Fig. 5.

1) STEP 01
Through the application, any person can search whether the corresponding cell phone number is enlisted in the suspected list.

2) STEP 02
If the cell phone user is already suspected by the application, the person will be instructed to answer some questionnaires which can be taken from [36, 37, 38]. The sample question set is shown in Table 1.

3) STEP 03
The answers will be verified and analyzed in real-time with a predefined answer set.

4) STEP 04
Finally, after the analysis, if the person is found to be possibly infected then the person will be instructed to go for a COVID-19 test.

These three phases are combined to form the proposed contact tracing method and the overall flow is depicted in Fig. 6.

TABLE I
SAMPLE QUESTIONNAIRE

| Sample Questions | User Options | |
|---|---|---|
| | Option-1 | Option-2 |
| 1. New or worsening cough | Yes | No |
| 2. Shortness of breath | Yes | No |
| 3. Sore throat | Yes | No |
| 4. Runny nose, Sneezing or nasal congestion | Yes | No |
| 5. Hoarse Voice | Yes | No |
| 6. Difficulty Swelling | Yes | No |
| 7. Nausea/vomiting/diarrhea/abdominal pain | Yes | No |
| 8. Unexpected fatigue | Yes | No |
| 9. Fever | Yes | No |

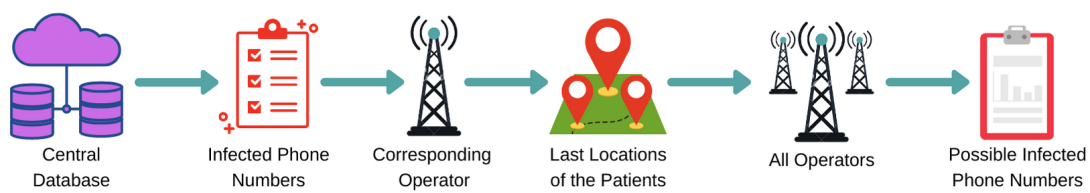

**FIGURE 4.** Flow diagram of Phase - II

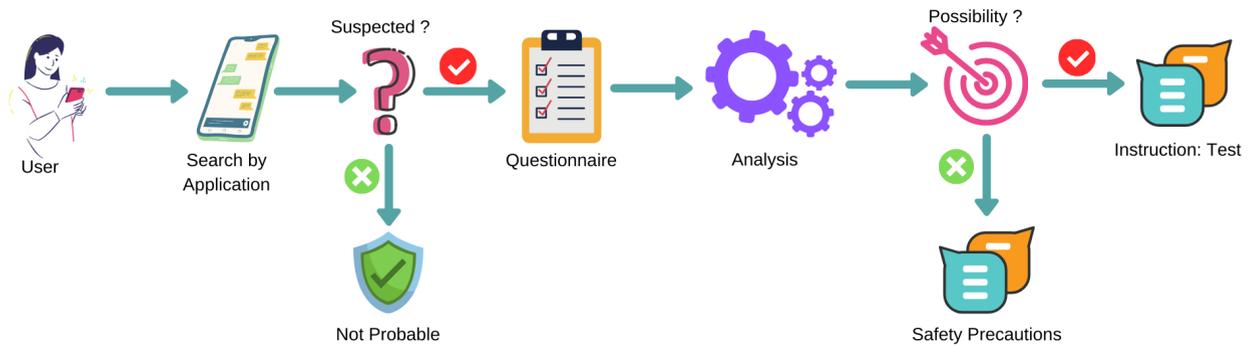

**FIGURE 5.** Flow diagram of Phase - III

## IV. COMPARISON WITH OTHER SOLUTIONS

The strongest aspect of the proposed contact tracing method is that it is not limited by advanced technological requirements (e.g. Smartphone, Bluetooth, Wi-Fi, NFC, and so on). The only requirement is that every individual has to carry the cell phone with an active sim card. With 5.112 billion unique mobile users, 67% penetration globally (even higher in urban areas where COVID-19 has the worst effect), mobile phone geolocation data seems to be the most viable means of contact tracing. Recently, a few articles have summarized different approaches proposed in the literature for minimizing the spread during this outbreak of COVID-19 [39], [40]. Most of the approaches are modern technology facilitated. The common approaches are broadcasting, selective broadcasting, unicasting, participatory sharing, private kit: safe paths, etc. Each of these approaches has its strengths and limitations. Raskar et al. have provided a detailed summary of the approaches named above [39]. We have reviewed the detailed comparison presented in [39] and analyzed different aspects of each method. The comparison of strengths and limitations of existing and the proposed methods is outlined in Table 2. The comparison factors are illustrated below as per the proposed model:

### A. ACCURACY
As the mobility information will be collected from the mobile operators, it is obvious that the data will be accurate. In the proposed model, we will not collect the locations verbally. As a result, we can say that the data will be precise and accurate.

### B. ADOPTION
There will be no adaptability issue as the person does not need to carry out any predefined activity. People will continue their day-to-day activities, and what everybody does most of the time, they will carry their cell phones.

### C. PRIVACY
The privacy issues are considered in the viewpoints of carriers, local businesses, users of the proposed model, and the non-users. These issues are discussed below:

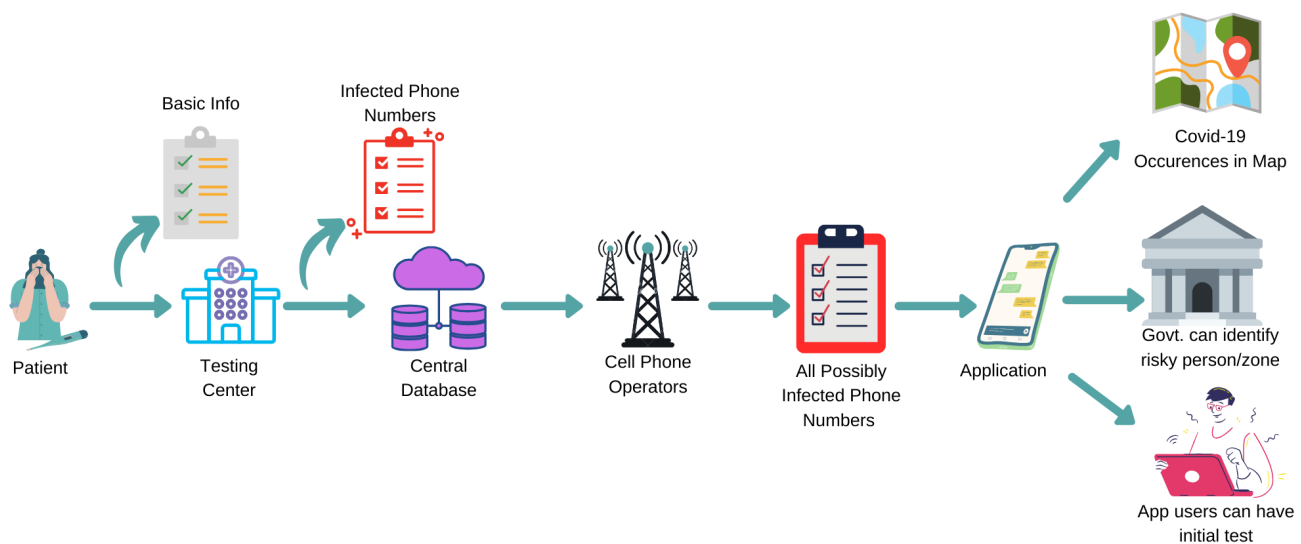

**FIGURE 6.** Overall Flow diagram of the proposed system

### 1) PRIVACY RISK FOR CARRIERS

There will be no privacy risks for the carriers as the personal information will not be publicized. Only the infected count will be made public via any map services (e.g. Google map). While identifying the adjacent persons of a particular cell phone user, the location information will not be made public also because the proposed model works in a centralized manner. The location data will be used only for predicting infection probability of the respective user.

### 2) PRIVACY RISK FOR LOCAL BUSINESS

There will be very low privacy risks for the local businesses as the locations where the carrier visited will not be made public.

### 3) PRIVACY RISK FOR USERS

The privacy of the users will be protected as the application will collect the location data from the sim card operator, not from the device location through Bluetooth or other means.

### 4) PRIVACY RISK FOR NON-USERS

There may be privacy violations for the non-users as the users and non-users are somehow connected via social relationships, but this is common for other contact tracing methods as well. When a person is diagnosed as COVID-19 positive, the family members or friends may endure the same unintended consequences of the event.

### D. CONSENT

The consent issues are considered in terms of the carriers, businesses and users of the proposed model. These issues are illustrated below:

### 1) CONSENT OF CARRIERS

The proposed model is considered to operate in such a way where the information about the patients will be recorded before the test of COVID-19. As a result, there will be no way to hide the details of the carrier afterward. Therefore, no consent will be needed as it is a casual and normal practice.

### 2) CONSENT OF LOCAL BUSINESSES

The consent of local businesses in the proposed model mainly depends on the Government policies.

### 3) CONSENT OF USERS

The consent of the users is required for the proposed model. During the registration, the user will be asked to provide the cell phone number. Even when the user wants to check if he/she made any contact with any COVID-19 patient, the cell phone number needs to be sent to the operator for a cross-check.

### E. SYSTEMATIC CHALLENGES

The systematic challenges are considered in terms of carriers, businesses, and users of the proposed model. The issues are described below:

### 1) MISINFORMATION

There is a very low risk of misinformation as the location data will not be taken from any user input. Rather the data will be collected from the sim card operator.

### 2) PANIC

The proposed model can reduce panic to some extent as the users of the model can check the initial status and also the intensity while staying at home.

TABLE II
STRENGTH AND LIMITATION ANALYSIS WITH OTHER PROPOSED METHODS

| Models | Major Strengths | Major Limitations |
|---|---|---|
| Broadcast | • No public adoption issues.<br>• User's privacy is protected.<br>• Mostly accessible by all. | • Significant privacy risk for COVID-19 carriers.<br>• Data accuracy issue.<br>• High risk of misinformation and mass panic. |
| Selective Broadcast | • Moderate privacy risk for COVID-19 carriers.<br>• Mass panic may be limited. | • Data accuracy issue.<br>• Risk of misinformation.<br>• Limited by technological requirements |
| Unicast | • Low risk of mass panic.<br>• Data accuracy is high. | • No privacy for the user<br>• Not helpful for the mass people. |
| Participatory | • User's privacy is protected from mass people. | • There can be fraudulent activity.<br>• Mass user adoption is low.<br>• Full consent is needed from the COVID-19 carrier. |
| Private Kit: Safe Path | • Overall accuracy is high.<br>• Moderate to low privacy risk for the COVID-19 carrier. | • Limited by technological requirements<br>• Full consent is required from the COVID-19 carrier. |
| Proposed Solution | • High accuracy of location data.<br>• Users' cell phones must not be smartphones rather any phone with an active sim card.<br>• To collect the data, no user application is needed.<br>• Less false negative issue. | • The user has to carry the cell phone. |

### 3) FRAUD AND ABUSE

It is expected that there will be no fraudulent activities as the application will not require any open connection (e.g. Bluetooth, NFC, etc.).

### 4) SECURITY OF INFORMATION

The proposed model is designed with a view in mind about the security concerns of the carrier's as well as the user's important information. As only the cell phone number, which is already a public entity, is needed to operate the proposed method, there is a very low chance of security holes.

### 5) EQUAL ACCESS
Equal access is the main strength of the proposed model as it is not limited by any technological requirement (e.g. smartphones, battery, certain OS, etc.) rather only a cell phone is needed.

### 6) SOCIOECONOMIC FACTORS
There are almost no bad impacts on the socioeconomic factors for the proposed model. Sometimes it is dependent on the government's practice if there is any.

Finally, the overall comparison of existing contact tracing approaches with the proposed approach is provided in Table II in terms of major strengths and limitations.

## V. EXPERIMENTAL RESULTS
In this section, we compare the proposed contact tracing method with existing technologies through numerical examples. To understand the real impact of our proposed system and to make a realistic comparison with the existing solutions based on BLE technology, we have generated the mimic of an actual contact tracing scenario based on some real statistical measures. In the scenario, we have used the Poisson Point Process (PPP) for a circular area of radius 1. We have scaled a distance of 100 meters to unity for ease of presentation. The percentages of smartphone users and any type of mobile phone users are considered as the density parameters of PPP for generating two separate sets of data.

We consider Bangladesh as a typical example of developing countries to show the practical effectiveness of the proposed approach. According to the Bangladesh Bureau of Statistics (BBS) and Bangladesh Statistics 2018, the total population of Bangladesh is 162.7 Million [41]. According to the Bangladesh Telecommunication Regulatory Commission (BTRC), the total number of mobile phone subscribers is 162.290 million [42] and the total number of mobile Internet subscribers is 93.101 million at the end of April 2020 [43]. As the total number of mobile phone subscribers is almost identical to the total population, we assume that almost 100% people have a mobile phone of some kind. Note that the official data of unique mobile users is not available during this analysis. We are considering the percentage of smartphone users based on the total number of mobile Internet subscribers. Thus, we can say that almost 58% of Bangladeshi people are using smartphones. Without loss of generality, we assume that people need smartphones to be able to use the BLE-based tracing app.

Now, we use the value of the density parameter for any type of mobile phone users as $\lambda = 100$ and for smartphone users as $\lambda = 58$ in Poisson Point Process. Based on this assumption, we simulate the data for both type of mobile phone users in Cartesian two-dimensional space. We have repeated the simulation four times so that we can make a reliable inference with a more accurate performance measure. In order to understand the dispersion of the points, we use scatter plots (Figs. 7 and 8) and covariance measures.

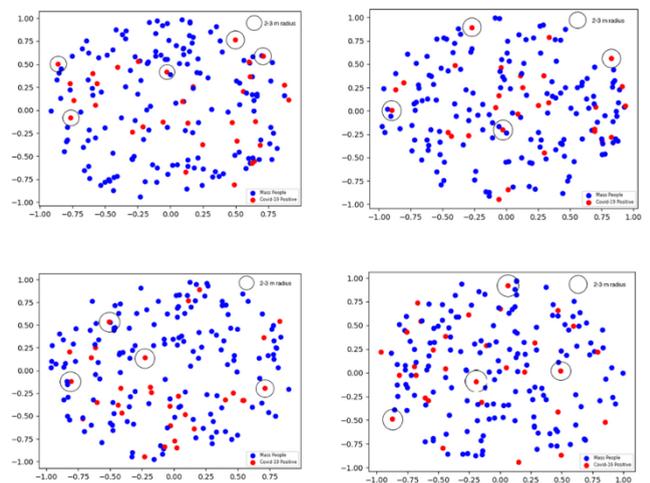

**FIGURE 7.** Scatter Plots (four trials) using Poisson Point Process ($\lambda = 58$).

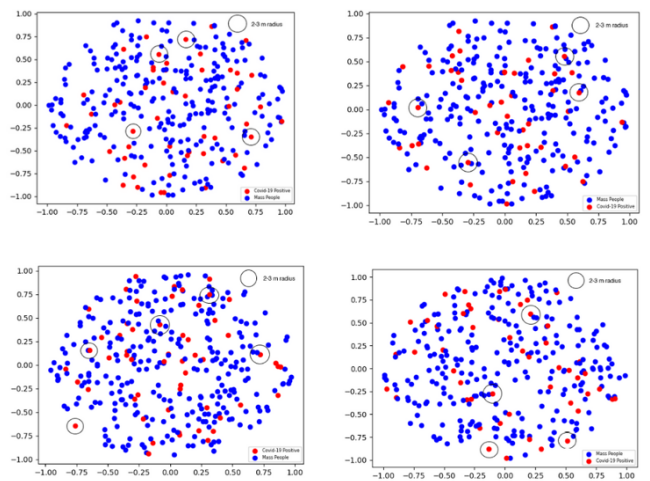

**FIGURE 8.** Scatter Plots (four trials) using Poisson Point Process ($\lambda = 100$).

TABLE III
COVARIANCE ANALYSIS OF THE GENERATED POINTS USING POISSON POINT PROCESS

| Trial | $\lambda = 58$ | $\lambda = 100$ |
|---|---|---|
| T-1 | 0.00634 | 0.002947 |
| T-2 | 0.01175 | 0.009763 |
| T-3 | 0.024061 | 0.00306 |
| T-4 | 0.00859 | 0.00582 |

It is apparent that the points in any type of mobile phone users (Fig. 8) are more condensed than the points simulated based on only smartphone users (Fig. 7). The covariance analysis in Table III is also supporting this observation to draw the same inference as the covariance is always smaller for the points in any type of mobile phone users (λ = 100) than smartphone users (λ = 58).

According to the official statistics, 18% of the people who were tested for COVID-19 in Bangladesh were found positive [44]. So, we have randomly selected 18% of the points as COVID-19 positive and marked with red dots in Fig. 7 and Fig. 8. Then, we have measured the Euclidian distance of every point with respect to the red points (likely COVID-19 positive). Finally, we have counted the points which were within a 3-meter radius of any of the Covid-19 positive points. Similarly, we count the same considering the 2-meter distancing strategy. The contact tracing performance comparison for a maximum of 3-meter distance is shown in Table IV. Here, we can see that for each of the four trials, the count of contact tracing using any type of mobile phone is higher than that of smartphone users. During the trials, the highest percentage of change with respect to smartphone users is 300% and the lowest is 100%. The average count of contact tracing for any type of mobile phone users and smartphone users is 20 and 7.75, respectively. These observations clearly show the effectiveness of the proposed tracing strategy.

TABLE IV
CONTACT TRACING COMPARISON FOR MAXIMUM 3-METER DISTANCE

| Trial | Contact Tracing Count | | Percentage of change with respect to the only Smartphone user |
|---|---|---|---|
| | For any type of Mobile Phone user | For Smartphone user only | |
| T-1 | 19 | 9 | 111.11 % |
| T-2 | 22 | 11 | 100 % |
| T-3 | 19 | 6 | 216.67 % |
| T-4 | 20 | 5 | 300 % |

The contact tracing comparison for a maximum of 2-meter distance is shown in Table V, where we again notice that for each trial, the count of contact tracing using any type of mobile phone is higher than that of smartphone users. During the trials, the highest percentage of change with respect to smartphone users is 266.67% and the lowest is 60%. The average count of contact tracing for any type of mobile phone users and smartphone users is 9.5 and 4.25, respectively.

TABLE V
CONTACT TRACING COMPARISON FOR MAXIMUM 2-METER DISTANCE

| Trial | Contact Tracing Count | | Percentage of change with respect to the only Smartphone user |
|---|---|---|---|
| | For any type of Mobile Phone user | For Smartphone user only | |
| T-1 | 9 | 5 | 80 % |
| T-2 | 10 | 4 | 150 % |
| T-3 | 8 | 5 | 60 % |
| T-4 | 11 | 3 | 266.67 % |

These two contact tracing comparisons are depicted graphically in Fig. 9 and Fig. 10.

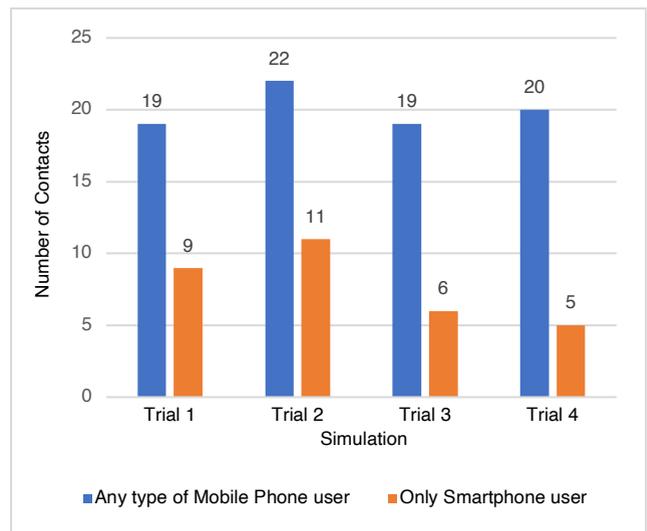

FIGURE 9. Pairwise contact tracing comparison considering the maximum 3-meter distance.

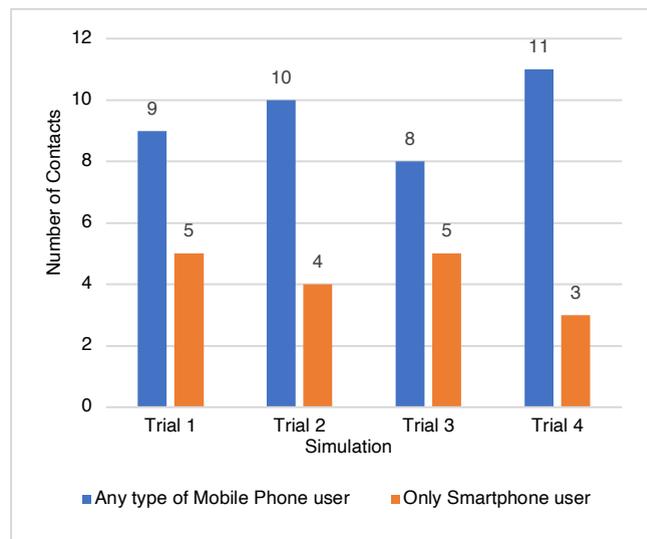

FIGURE 10. Pairwise contact tracing comparison considering the maximum 2-meter distance.

Next, we extend the performance comparison to the case of two other countries, e.g. India and South Korea because South Korea is ranked at the top of the list for smartphone usage (95%) and India at the last (24%) based on the total population [45]. In India, 24% of the total population use smartphones, 40% people use mobile phones that are not smartphones and 35% people do not use any mobile phone at all. Therefore, we simulate the data for India with the density parameter, λ = 24 for smartphone-only users and λ = 64 for any kind of mobile phone users. On the other hand, in South Korea, 95% people use smartphones and 5% use a mobile phone that are not smartphones. So, for South Korea, we consider the density parameter, λ = 95 for smartphone-only users and λ = 100 for any kind of mobile phone users. According to COVID-19 case statistics for both the countries, it is found that in India, 4.16 % of people were found positive after test [46] and it is 1.06 % in South Korea [47]. We have used these percentages to consider the COVID-19 positive points. The country-wise contact tracing performance of existing and proposed schemes is shown in Fig. 11.

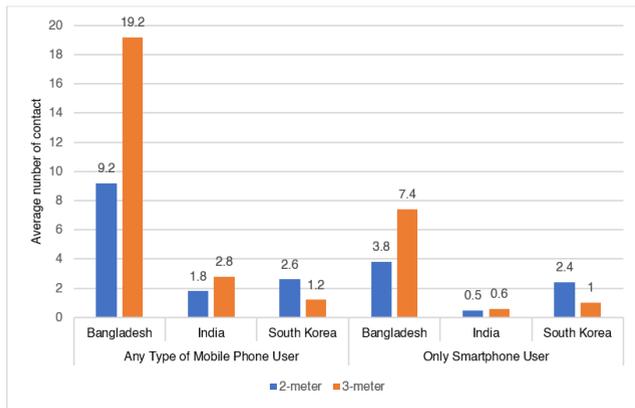

**FIGURE 11.** Country wise average contact tracing count comparison.

From Fig.11, it is noticeable that the count of contact tracing is higher for most of the time when any type of mobile phone users' geolocation is considered than smartphone-based apps. The percentage of changes with respect to smartphone users for each of the countries is also given in Fig. 12.

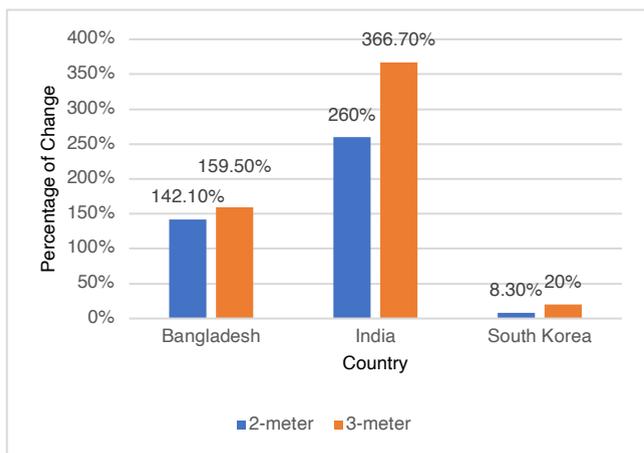

**FIGURE 12.** Country wise comparison on percentage of change with respect to only smartphone users

## VI. CONCLUSIONS

To slow down the spreading of the deadly virus termed as the novel Coronavirus, there is no alternative to finding out the infected persons as well as those who came into close contact with an infected person, then taking proper measures. Considering this fundamental approach, many countries have already developed contact tracing applications which are showing promising results but with significant privacy concerns. In this paper, we have addressed the privacy issue by avoiding any smartphone-based apps for contact tracing through wireless connectivity (e.g. Bluetooth, Wi-Fi, NFC, etc.). The proposed model uses mobile users' geolocation data directly from the mobile operators. By doing so the overall contact tracing performance improves significantly while preserving users' privacy.

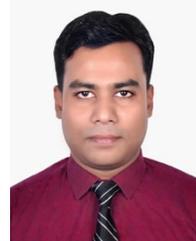

**MD. T. RAHMAN** was born in Meherpur, Bangladesh. He received the B.Sc. (Hons) degree from Jahangirnagar University, Bangladesh, in 2013 and the M.S. degree from the same university in 2015, both in Information Technology.

From 2015 to 2019, he was working as a Lecturer in the Department of Computer Science and Engineering, Daffodil International University, Dhaka, Bangladesh. He is currently working as a Lecturer in the Department of Information and Communication Technology, Mawlana Bhashani Science and Technology University, Bangladesh. His research interests include Wireless Communication & Internet Security Concerns.

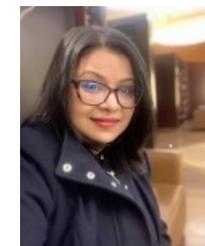

**RISALA T. KHAN** was born in Dhaka, Bangladesh. She has completed her B.Sc. (Hons) in Computer Science and Engineering from Jahangirnagar University, Savar, Dhaka in 2003, M.Sc. in Computer Science and Engineering in 2005. She has completed her Ph.D. from the same University in the field of Cognitive Radio System.

She worked as a Lecturer in the Department of Computer Science and Engineering, Daffodil International University, Dhaka. She is now working as a Professor at Institute of Information Technology, Jahangirnagar University, Savar, Dhaka, Bangladesh. Her research field is wireless communications, network traffic and network security.

Dr. Khan is a member of IEEE and also acting as counselor of IEEE WIE Affinity Group JU SB.

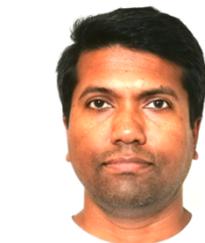

**MUHAMMAD R. A. KHANDAKER** (S'10–M'13–SM'18) received his PhD degree from Curtin University, Australia. He is currently an Assistant Professor in the School of Engineering and Physical Sciences at Heriot-Watt University. Before joining Heriot-Watt, he worked as a Postdoctoral Research Fellow at University College London, UK, (July 2013 - June 2018). He is an Associate Editor for the IEEE Wireless Communications Letters, IEEE COMMUNICATIONS LETTERS, the IEEE ACCESS and the EURASIP JOURNAL ON WIRELESS COMMUNICATIONS AND NETWORKING.

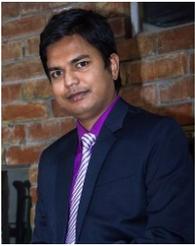

**MD. S. A. SALAN** was born in Panchagarh, Bangladesh. He has completed his B.Sc. (Hons) in Statistics from Jahangirnagar University, Savar, Dhaka in 2015, M.Sc. in Statistics in 2016 from the same University. He is now working as an Assistant Professor at the Department of Statistics, Mawlana Bhashani Science and Technology University, Tangail, Bangladesh. Before that, he worked as a permanent faculty of Statistics at Primeasia University and Daffodil International University. His research field is Statistical Data Modeling with Simulation, Machine Learning and Time Series Analysis.